\documentclass[%
 aip,
 amsmath,amssymb,
     reprint,%
]{revtex4-1}
\usepackage[utf8x]{inputenc}
\usepackage[pdftex]{graphicx}
\usepackage[squaren]{SIunits}

\begin{document}
\preprint{AIP/123-QED}
\title{Nonuniform current and spin accumulation in a \unit{1}{\micro\metre} thick n-GaAs channel}
\author{B. Endres, M. Ciorga, R. Wagner, S. Ringer, M. Utz, D. Bougeard, D. Weiss, C. H. Back and G. Bayreuther}
\affiliation{Institut für Experimentelle und Angewandte Physik, Universität Regensburg, Germany}
\begin{abstract}
The spin accumulation in an n-GaAs channel produced by spin extraction into a (Ga,Mn)As contact is measured by cross-sectional imaging of the spin polarization in GaAs. The spin polarization is observed in a 1 µm thick n-GaAs channel with the maximum polarization near the contact edge opposite to the maximum current density. The one-dimensional model of electron drift and spin diffusion frequently used cannot explain this observation. It also leads to incorrect spin lifetimes from Hanle curves with a strong bias and distance dependence. Numerical simulations based on a two-dimensional drift-diffusion model, however, reproduce the observed spin distribution quite well and lead to realistic spin lifetimes.
\end{abstract}
\maketitle
Spin injection from a ferromagnetic contact into a semiconductor is a fundamental prerequisite for many spintronic devices \cite{RevModPhys.76.323}$^,$ \cite{2007AcPSl..57..565F}. Moreover, a reasonably large spin lifetime in the semiconductor is necessary for any useful application. The value of the spin lifetime is often extracted from Hanle-curves obtained from non-local voltage measurements or by optical means \cite{PhysRevB.79.165321}$^,$ \cite{2007NatPh...3..197L}$^,$ \cite{1367-2630-9-9-347}$^,$ \cite{10.1063/1.3097012} using a one-dimensional spin drift-diffusion model.
Parasitic contributions such as dynamic nuclear polarization at low temperatures have to be avoided by fast periodic magnetization reversal of the injecting contact \cite{10.1063/1.3590726}. Here we show that, in addition, the electric field near the injecting contact strongly influences the spin density distribution around it. As a result a one-dimensional picture of drift and diffusion along the channel leads to incorrect spin lifetime values.
This becomes obvious by comparing fitting results of Hanle data from a one- and a two-dimensional model.
For our studies we employ a cross-sectional imaging method that allows to probe the two-dimensional spin polarization distribution even below the contacts \cite{2007kotissek}$^,$ \cite{2011JAP...109gC505E} and compare the experimental results to two-dimensional drift-diffusion simulations.

\begin{figure}[t]
\centering
 \includegraphics[width=0.4\textwidth]{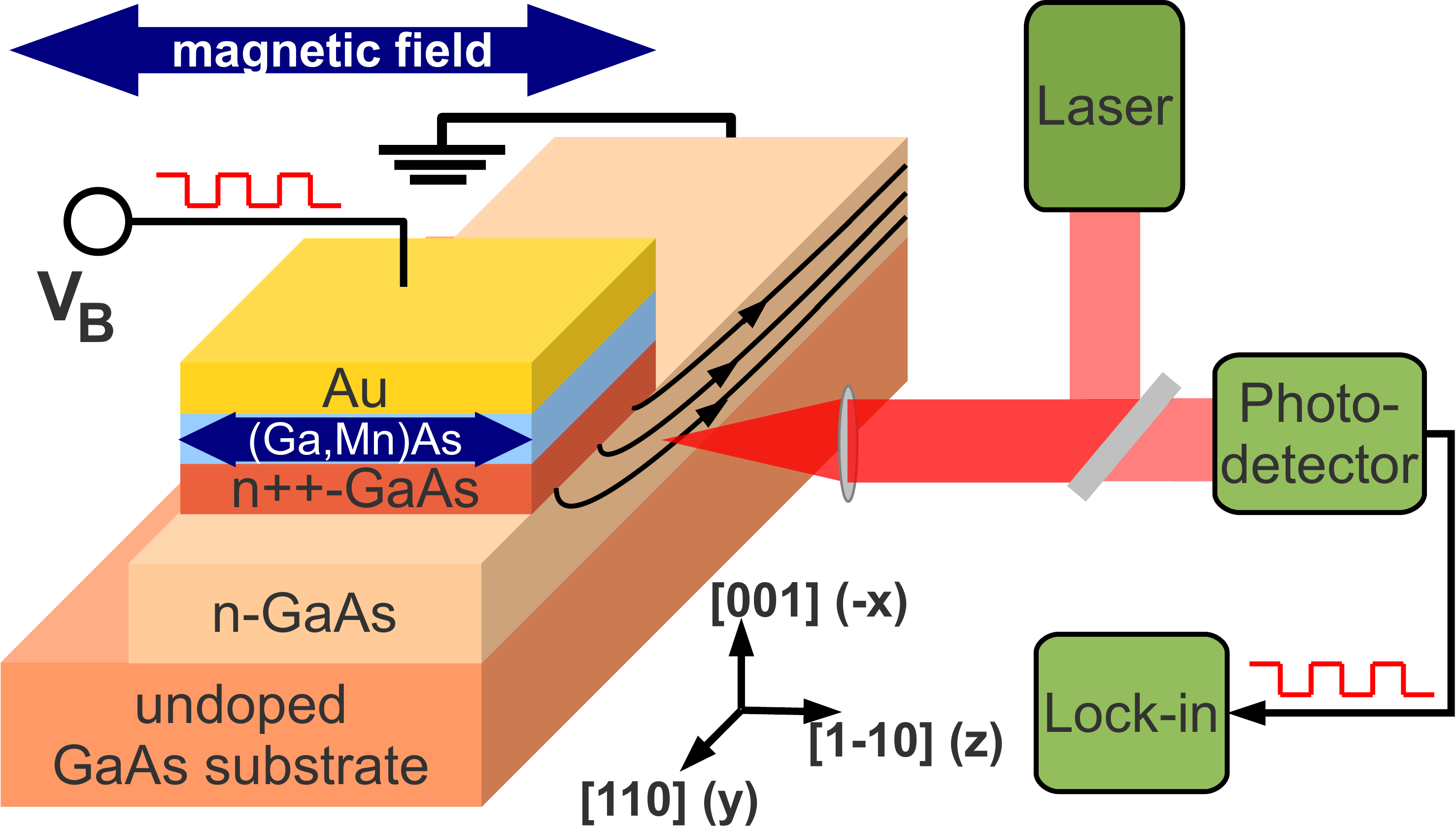}
\caption{Geometry of the sample and measurement principle.}
\label{setup}
\end{figure}
The measurement principle and sample geometry are sketched in Fig.~\ref{setup}.
The layers are grown by molecular beam epitaxy on a semi-insulating GaAs(001) substrate.
The \unit{1}{\micro\metre} thick n-GaAs channel for the electron transport and spin accumulation has a doping density of \unit{2.7\cdot10^{16}}{cm^{-3}}.
Spin injecting contacts are formed by a 50 nm thick layer of the diluted magnetic semiconductor $\textrm{Ga}_{95}\textrm{Mn}_{5}\textrm{As}$, a 8 nm thick layer of n$^{++}$-GaAs doped up to \unit{5\cdot10^{18}}{cm^{-3}} and a 15 nm thick \mbox{n$^{++}\rightarrow$ n} transition layer. Together these layers form a spin Esaki diode that enables tunneling of spin-polarized electrons between (Ga,Mn)As and the GaAs channel.
The detailed layer stack and the contact patterning are described elsewhere \cite{epub21452}.
The fabricated lateral device has a \unit{50}{\micro\metre} wide mesa channel, one \unit{4}{\micro\metre}$\times$\unit{50}{\micro\metre} injecting contact and two large reference contacts located at a distance of \unit{300}{\micro\metre} to the Ga(Mn,As) contact.
Finally, the sample is cleaved along the [110] direction across the mesa channel and the ferromagnetic pad, reducing the contact area to \unit{4}{\micro\metre}$\times$\unit{40}{\micro\metre}. The cleaving process furthermore exposes the GaAs(1-10) surface and enables a direct optical access to the n-GaAs channel (see Fig.~\ref{setup}). For the optical measurements the sample is mounted in a He flow cryostat. The cryostat itself is mounted on top of a nano positioner enabling two-dimensional scans in the xy-plane. The z component of the electron spin polarization (that is, the component along [1-10], which is a magnetic easy axis of the Ga(Mn,As) contact) in the n-GaAs channel is detected via the polar magneto-optical Kerr effect (pMOKE) with a spot size of \unit{1}{\micro\metre}. The photon energy of the linearly polarized laser beam was chosen close to the band-gap of GaAs ($\lambda=816$ nm at 9 K), where the specific Kerr rotation shows a maximum (not shown).
A square-wave bias voltage alternating between zero and $V_B$ is applied between the contacts and the Kerr rotation is detected synchronously with balanced photo-receivers and a lock-in technique. This ensures that the (quasi-static) magnetization of the ferromagnetic contacts does not contribute to the Kerr signal \cite{2007kotissek}$^,$ \cite{2011JAP...109gC505E}.

\begin{figure}[tb]
\centering
 \includegraphics[width=0.47\textwidth]{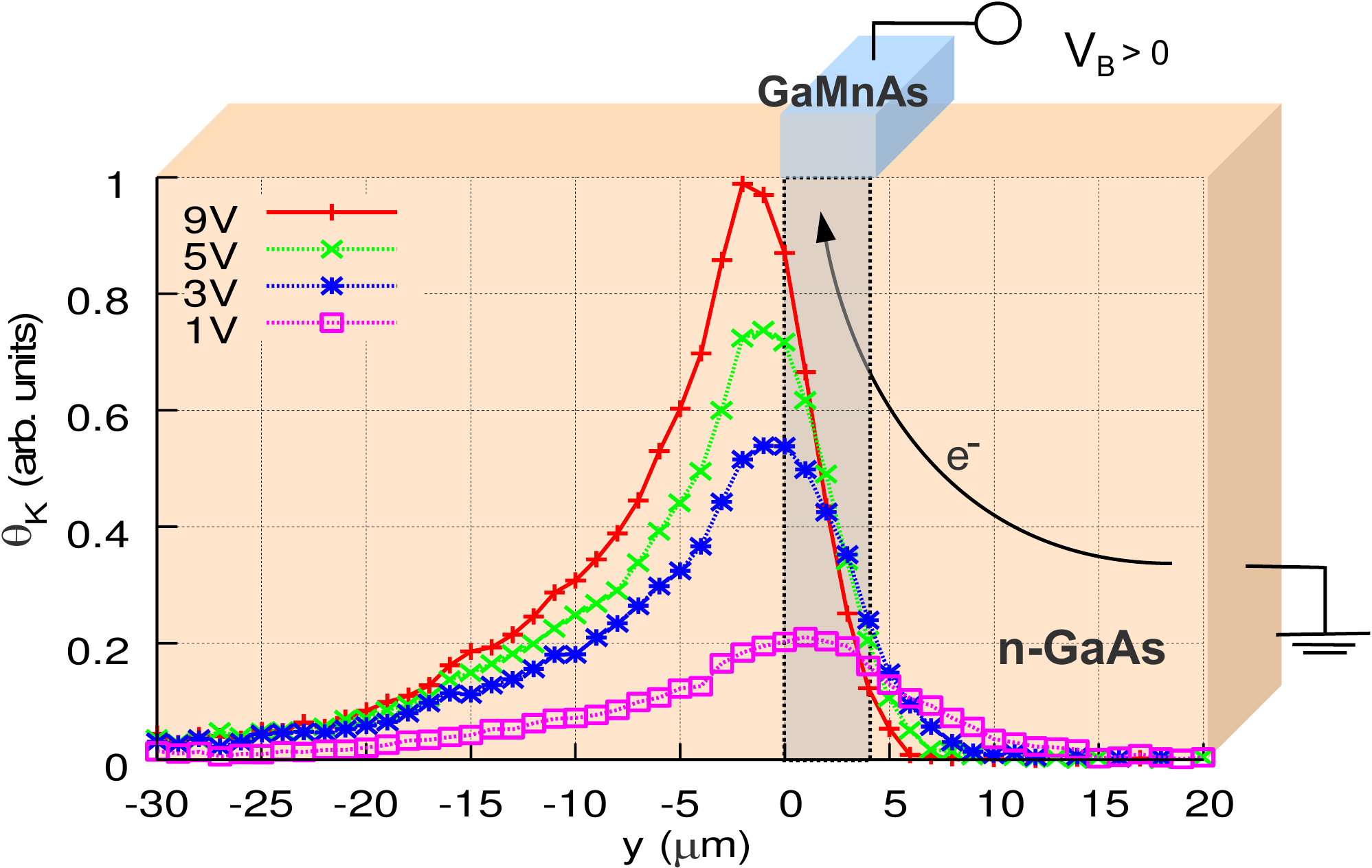}
\caption{Kerr rotation along the n-GaAs channel for various applied voltages, representing the local spin polarization.}
\label{2dscan} 
\end{figure}

Fig.~\ref{2dscan} shows one-dimensional scans of the Kerr rotation along the n-GaAs channel for different applied positive voltages between the (Ga,Mn)As contact in the middle and the reference contact at the right hand side. As illustrated in Fig.~\ref{2dscan}, this denotes that  unpolarized electrons in the channel are accelerated to the left and the spin accumulation is generated by an electron flow into the (Ga,Mn)As contact. Thus, spins are extracted from the n-GaAs layer. To eliminate any electro-optic background, the measurements are performed in remanence after saturation along [1-10] and [-110], respectively and the difference between both remanent values is used as a measure of the spin polarization in the GaAs layer.

The observed spin density distribution in Fig.~\ref{2dscan} shows an unexpected maximum spin polarization that is located near the contact edge \textbf{opposite} to the maximum charge current. This peak shifts even further to the left with increasing bias voltage. On the left hand side of the contact the spin density distribution should be solely driven by diffusion. However, the measured decay of the spin polarization on the diffusion side cannot be described by a single exponential decay as expected (not shown). Depending on the bias voltage, a faster decay close to the contact edge is observed. Since the decay constant defines the spin diffusion length $L_s=\sqrt{D_s\cdot \tau_s}~,$ either the spin diffusivity or the spin lifetime must be reduced in this area. A small bias dependence is still visible when fitting the curves from y-position \unit{-30}{\micro\metre} to \unit{-7}{\micro\metre}, where a spin diffusion length of \unit{8.2}{\micro\metre} was extracted for the largest bias (9V / \unit{750}{\micro\ampere}) and \unit{9.2}{\micro\metre} for the smallest bias (1V / \unit{25}{\micro\ampere}). On the right hand side of the contact, the drift side, the spin distribution can be described by a superposition of drift and diffusion. This explains the shortened effective decay length with increasing bias voltage, but it does not explain the decay of the spin polarization that already shows up beneath the contact area. As a verification of this spin density distribution, we also perform non-local voltage measurements on a similar sample with additional (Ga,Mn)As contacts on the drift and diffusion side. The result from the electrical detection experiment shows a good agreement with the optical data (not shown).

\begin{figure}[t]
\centering
 \includegraphics[width=0.43\textwidth]{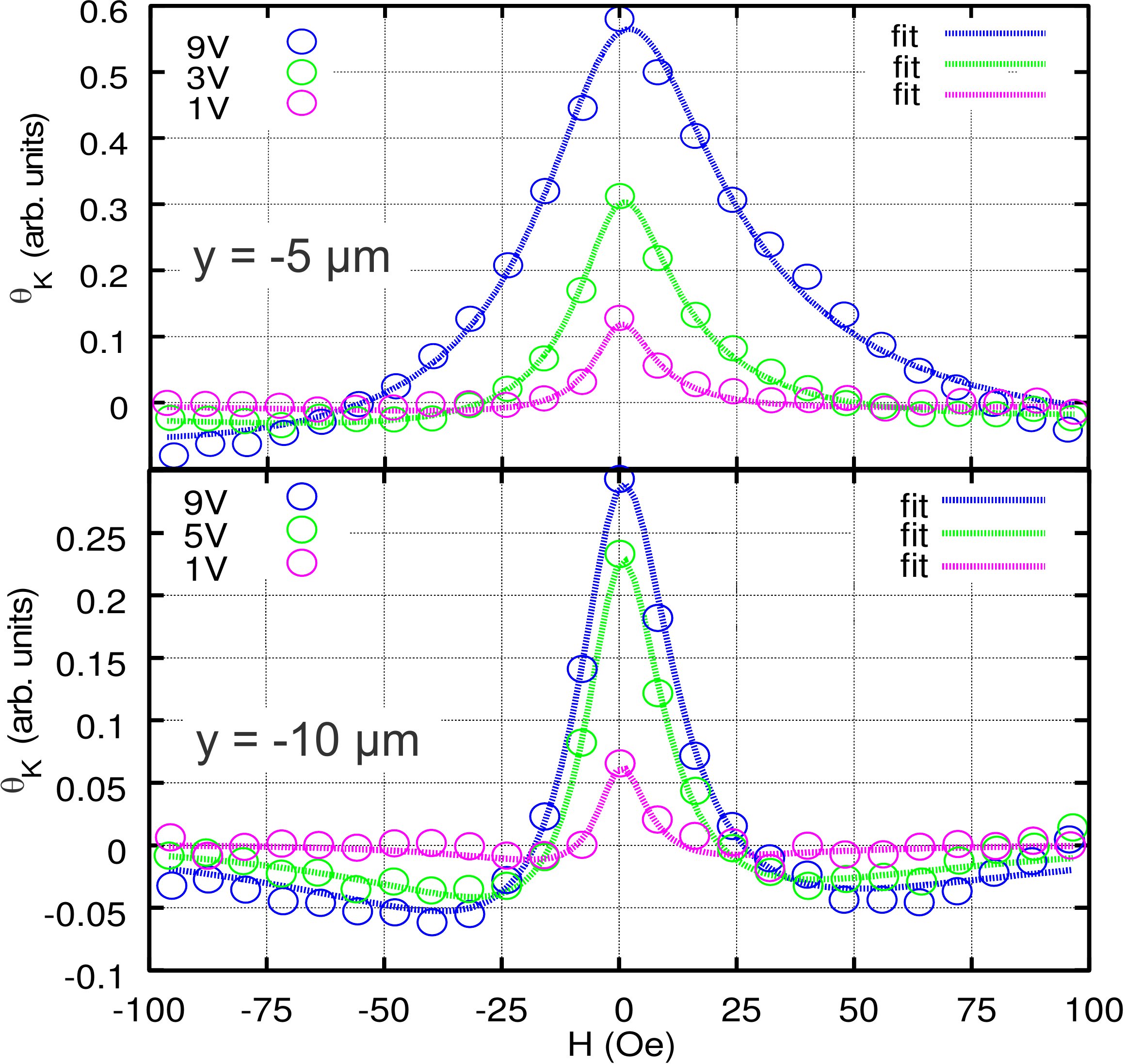}
\caption{Hanle-curves for different applied bias voltages, measured at y-position \unit{-5}{\micro\metre} (top) and \unit{-10}{\micro\metre} (bottom). Fits using a one-dimensional model yield different values of the spin lifetime which are shown in TABLE I.} 
\label{hanle}
\end{figure}
For a better understanding of the observed spin density distribution and its bias dependence we also perform Hanle-measurements at y-position \unit{-5}{\micro\metre} and \unit{-10}{\micro\metre} for different applied voltages, illustrated in Fig.~\ref{hanle}. The curves are fitted using a one-dimensional function \cite{suppl}. Assuming pure spin diffusion from the contact area to the measurement position, the only free parameter left is the spin lifetime, as the spin diffusion length is well known from the previously shown results (see Fig.~\ref{2dscan}). However, this approach leads to spin lifetimes depending significantly on the applied bias voltage and the injection-detection distance (see Fig.~\ref{hanle}), with values changing from 14 ns for a large voltage and small distance to 46 ns for a small voltage and large distance.
The bias dependence is less pronounced when measuring further away from the contact, but still a factor of two difference is observed. However, even the 1V-data with a current of \unit{25}{\micro\ampere} yield different values at both positions (42 and 46 ns). 

To shed light on this behavior we perform numerical simulations of the two-dimensional electron drift and spin diffusion. The simulation is done in two steps: first the current density distribution is simulated in the n-GaAs channel that in principle depends on the interface resistance of the Esaki diode and the channel resistance. Both values are determined by electrical measurements, where the voltage drop across the interface is detected via the second reference contact. At 9~K, the channel's conductivity is 1500~S/m, resulting in an electron mobility $\mu$ of 0.35~$m^2/(s\cdot V)$ in the n-GaAs channel. In the simulation, the Esaki diode is represented by a 10~nm thick layer with a conductivity of 0.025~S/m, matching with the experimentally determined resistance for this bias voltage. Based on the result for the current density distribution, the spin density distribution is calculated by assuming a spin current across the interface that is proportional to the current density. The generated spin accumulation then spreads due to diffusion and the previously calculated drift in this sample geometry until a steady state is found. A derivation of the used differential equations and boundary conditions can be found in Ref.~2. The spin diffusivity in the simulation is defined by the spin diffusion length, extracted from the measurements shown in Fig.~\ref{2dscan} and the spin lifetime (39ns), estimated by Hanle-measurements (see below).

Fig.~\ref{comsol} illustrates the simulated spin density distribution for a large positive applied bias voltage at the (Ga,Mn)As contact, which reproduces the observed spin density distribution quite well: the pronounced nonuniform electron drift beneath the contact area shifts the polarization peak towards the diffusion side. This shift increases with applied bias voltage similar to our experimental results (see Fig.~\ref{2dscan}). The simulation also directly explains the extremely fast decay of the spin polarization on the right hand side of the contact, where the decay already begins beneath the contact area. 
\begin{figure}[t]
\centering
 \includegraphics[width=0.45\textwidth]{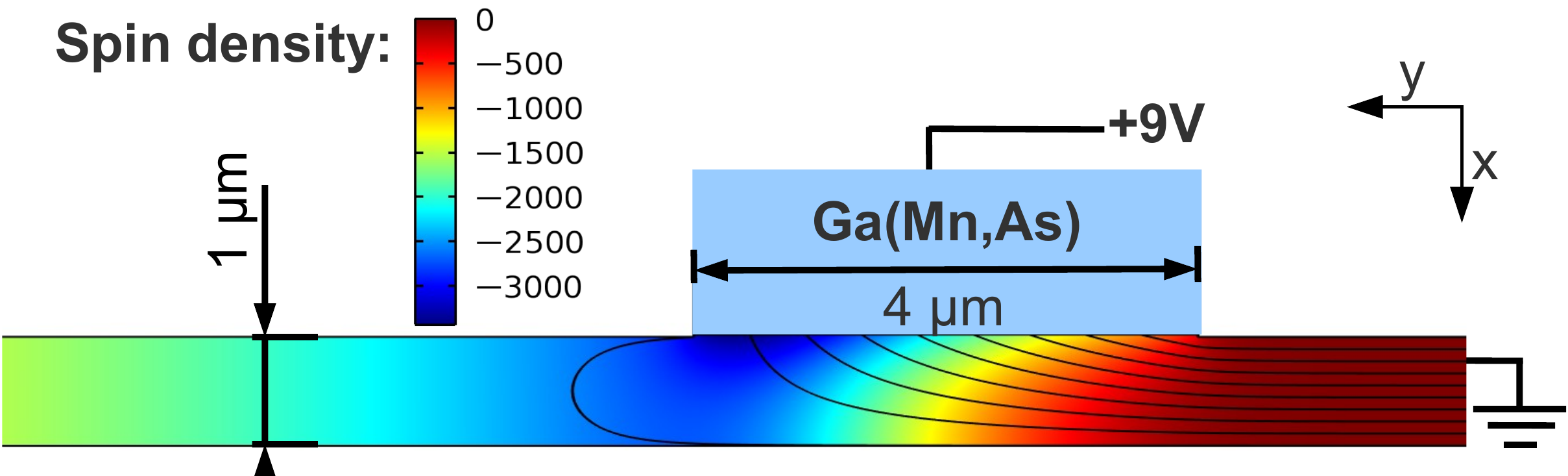}
\caption{Two-dimensional drift-diffusion simulation of the spin accumulation in a \unit{1}{\micro\metre} thick n-GaAs channel. Black lines represent the local current direction.} 
\label{comsol}
\end{figure}

An important point to note is the nonuniform current density on the diffusion side and underneath the contact area. As illustrated in Fig.~\ref{comsol} by the black lines representing the local current direction, the n-GaAs channel on the diffusion side can be divided into a bottom channel where electrons are flowing to the left away from the contact and a top channel with a back-flow towards the contact. This nonuniform current density decays rapidly with increasing distance but cannot be neglected within the first micrometer on the diffusion side. Due to this inhomogeneous current density it is not sufficient to describe the electron spins solely by diffusion on this side of the channel, which was done before when fitting the Hanle-curves. In particular, the electric field beneath the contact area generates a positive drift towards the measurement position. This additional drift widens the Hanle-curves and therefore the fit provides lower spin lifetimes. 

This effect becomes apparent when the Hanle data is fitted with our two-dimensional model where the bias dependent spin density distribution and the nonuniform drift is taken into account \cite{suppl}:
\begin{table}[h]
\begin{center}
\begin{tabular}{llllllll}
 & &\vline ~1 V & 3 V & 5 V & 7 V & 9 V\\
\hline
\textit{y=-5 µm}&\textit{1D} &\vline ~\textit{42 ns} & \textit{25 ns} & \textit{20 ns} & \textit{16 ns} & \textit{14 ns}\\
y=-5 µm&2D &\vline ~46 ns & 47 ns & 46 ns & 41 ns & 38 ns\\
\textit{y=-10 µm}&\textit{1D} &\vline ~\textit{46 ns} &  & \textit{29 ns} &  & \textit{23 ns}\\
y=-10 µm&2D &\vline ~53 ns &  & 53 ns &  & 47 ns
\end{tabular}
\caption{Spin lifetime from numerical simulations based on 1D and 2D drift-diffusion models at two different distances, y, from the contact edge and different bias voltages \cite{suppl}.}
\end{center}
\end{table}

A spin lifetime of 38 ns is extracted for the 9 V data in contrast to 14 ns for the one-dimensional model. Thus, the strong bias dependence of the spin lifetime from the one-dimensional model can be primarily attributed to the neglect of the nonuniform current. However, a variation of the spin lifetime with bias voltage and contact distance of about 20\% remains.

The dominant spin relaxation mechanism for n-doped GaAs above the metal-insulator transition is the Dyakonov-Perel mechanism \cite{PhysRevB.66.245204}. This mechanism strongly depends on the electron energy in the conduction band, for instance on the doping density or the temperature (in case of charged impurity scattering $\tau_s \varpropto T^{-{3/2}}$) \cite{RevModPhys.76.323}.
According to this formula, a temperature increase up to 2~K would be sufficient to explain the observed behavior.
Besides the temperature, also the electric field has an influence on the spin lifetime, as has already been shown for optically pumped spins in 2006 \cite{0295-5075-75-4-597}$^,$ \cite{10.1063/1.2345608}. The reason is the increased energy of the electrons when accelerated in electric fields leading to an increased momentum relaxation time $\tau_p$ and thus to a more efficient spin relaxation due to the Dyakonov-Perel mechanism \cite{RevModPhys.76.323}. A detailed calculation of this effect was reported by Beck et al. \cite{0295-5075-75-4-597}

From the one-dimensional scans and the simulation it is evident that electric fields are present around the contact area. Even for the 1V data, the electric field in the n-GaAs channel is in the order of 10~V/cm, that, according to Refs.~12 and 13 should already reduce the spin lifetime. Since strong electric fields are only located beneath the contact area and the electric field intensity decreases rapidly on the diffusion side, larger spin lifetimes and a less pronounced bias dependence are observed when measuring further away from the contact.

In summary, it is shown that a nonuniform current density in the n-GaAs channel affects the spin density distribution. The pronounced electric field beneath the contact area generates a drift towards the diffusion side and therefore widens Hanle-curves when measured on this side of the contact. As a consequence, extracting the spin lifetime with a one-dimensional model assuming pure spin diffusion leads to a strong bias and contact distance dependence. Hence a two-dimensional model has to be used to yield correct spin lifetimes. The remaining variations with distance and bias may be due to Joule heating or the presence of electric fields around the contact area.
As a final note, the nonuniform current density might also be the origin of offset-signals in non-local voltage measurements and directly shows that the non-local voltage is influenced by resistance changes of the injector contact.

Financial support by the Deutsche Forschungsgemeinschaft (SFB 689) is gratefully acknowledged.
%
\end{document}